\documentclass[preprint,nofootinbib]{revtex4}

\usepackage{graphicx}

\begin{document}

\title{Evolution of density perturbations in decaying vacuum
cosmology: The case of non-zero perturbations in the cosmological
term}

\author{H. A. Borges$^1$\footnote{E-mail address: humberto@ufba.br},
S. Carneiro$^{1,2}$\footnote{ICTP Associate Member. E-mail address: saulo.carneiro@pq.cnpq.br},
J. C. Fabris$^{3,4}$\footnote{E-mail address: fabris@pq.cnpq.br}}

\affiliation{$^1$Instituto de F\'{\i}sica, Universidade Federal da
Bahia, Salvador, BA, Brazil\\$^2$Astronomy Unit, School of
Mathematical Sciences, Queen Mary, University of London, London,
UK\\$^3$Departamento de F\'{\i}sica, Universidade Federal do
Esp\'{\i}rito Santo, Vit\'oria, ES, Brazil\\$^4$Institut
d'Astrophysique de Paris, Paris, France}

\begin{abstract}
We extend the results of a previous paper where a model of
interacting dark energy, with a cosmological term decaying linearly
with the Hubble parameter, is tested against the observed mass power
spectrum. In spite of the agreement with observations of type Ia
supernovas, baryonic acoustic oscillations and the cosmic microwave
background, we had shown previously that no good concordance is
achieved if we include the mass power spectrum. However, our
analysis was based on the {\it ad hoc} assumption that the
interacting cosmological term is strictly homogeneous. Now we
perform a more complete analysis, by perturbing such a term.
Although our conclusions are still based on a particular, scale
invariant choice of the primordial spectrum of dark energy
perturbations, we show that a cosmological term decaying linearly
with the Hubble parameter is indeed disfavored as compared to the
standard model.
\end{abstract}

\maketitle

\section{Introduction}

The elucidation of the nature of dark energy is one of the most
important challenges of modern cosmology, requiring at present the
attention of theoreticians and observational teams. From the
theoretical viewpoint, a crucial problem is to understand the role
of vacuum in the cosmological scenario, its possible relation with
dark energy and, in this case, why its observed density is so small
as compared to the value theoretically expected by quantum field
theories \cite{Pad}.

Among the different approaches to this problem, one can find the
suggestion that dark energy is the manifestation of quantum vacuum
in the curved space-time, and that its density depends on the
curvature, decaying from a huge initial value as space expands. As
the total energy must be conserved, the vacuum decay is concomitant
with matter production, a general feature of this kind of models
and, more generally, of interacting dark energy models
\cite{Ozer,Horvat,SC,Grande,Holografia,Zimdahl,Ma,Fabris}.

However, it is difficult to derive the vacuum contribution in the
expanding background, and some phenomenological approaches are
needed in order to implement the above idea. A thermodynamical
analysis in de Sitter space-time, in line with the holographic
conjecture, has suggested a particular dependence of the vacuum
density $\Lambda$ on the Hubble parameter $H$, which is a good
approximation in expanding, quasi-de Sitter space-times \cite{SC2}.

Such a dependence was obtained by noting that a free particle in the
de Sitter background presents, superposed to its normal modes, a
thermal motion with a characteristic temperature equals to $H$,
which is an expression of the known association of a temperature $H$
to the de Sitter horizon. On the other hand, if we regularize the
vacuum energy density in the flat space-time by postulating a
thermal distribution of the vacuum fluctuation modes at a
temperature $m$ (which is equivalent to impose a superior cutoff $m$
on the modes frequencies), we obtain $\Lambda_0 \approx m^4$. Now,
if in de Sitter space-time we shift this vacuum temperature from $m$
to $m+H$ we obtain, after subtracting the flat space-time
contribution, the ansatz $\Lambda \approx (m+H)^4 - m^4$.

This is a phenomenological ansatz, in the sense that there is still
no rigorous derivation of it in the realm of quantum field theories
in curved space-time. Other interesting ansatzen have been proposed
as well, also on phenomenological or semi-phenomenological basis
(see, for example, \cite{Grande,Holografia,Fabris}). In this context
the comparison with observations is relevant, since it can rule out
or constrain some models while we do not have a rigorous theoretical
answer to the problem.

If we use for the cutoff $m$ the energy scale of the QCD phase
transition (the latest known cosmological vacuum transition), in the
limit of very early times we have $H>>m$, and the cosmological term
is proportional to $H^4$. This leads to a non-singular inflationary
solution, with an initial quasi-de Sitter phase giving origin to a
radiation-dominated universe, with a matter content generated at the
expenses of vacuum energy \cite{SC2}.

For late times all depends on the masses of the produced particles.
The time-energy uncertainty relation suggests that
particles of mass $M$ can only be produced if $H > M$. Hence,
massive particles as the baryonic matter, axions and supersymmetric
candidates for dark matter should stop to be produced at the time of
electro-weak phase transition (whereas the production of photons and
massless neutrinos must be forbidden by some selection rule,
otherwise our universe would be completely different at present).
Therefore, if no other particle is taken into account, for late
times we have a genuine cosmological constant, like in the standard
$\Lambda$CDM model. However, if we consider the possibility of very
light dark particles (as massive gravitons, for example), the vacuum
decay could still happen. Since in this limit we have $H<<m$, now the vacuum density varies
linearly with $H$, that is, $\Lambda \approx m^3H$
\cite{SC2}. On the other hand, as in the present universe the cosmological term is
dominant, the Friedmann equation gives $\Lambda \approx H^2$. Hence, we have $H \approx m^3$ and
$\Lambda \approx m^6$. With $m$ of the order of the energy scale of
the QCD chiral phase transition (approximately the pion mass), these
expressions lead to numerical coincidences (in orders of magnitude)
with the observed current values of $H$ and $\Lambda$.

In this case, the resulting scenario is similar to the standard
one, with the radiation phase followed by a long matter era, with
the cosmological term dominating for large times \cite{Borges}. A
detailed analysis of the redshift-distance relation for type Ia
supernovas, the baryonic acoustic oscillations (BAO) and the
position of the first peak in the spectrum of anisotropies of the
cosmic microwave background (CMB) has shown a very good concordance
\cite{Jailson2}, with present values for the Hubble parameter, the
relative matter density and the universe age inside the limits
imposed by other, non-cosmological observations.

Nevertheless, a late-time matter production may dilute the density
contrast during the process of structure formation \cite{Reuven}. Therefore, the
study of the evolution of perturbations and the comparison of the
model predictions with the observed mass power spectrum is also
necessary. In a previous paper \cite{I} we have shown that, in the
case of a strictly homogeneous cosmological term (which means that
matter production is homogeneous as well), the matter contrast is
indeed suppressed at late times, after achieving a maximum near the
present epoch. This would be a potential explanation for the cosmic
coincidence, but it also leads to a suppression in the power
spectrum. As a consequence, a good accordance with observations is
only possible with a relative matter density above the concordance
value obtained from the joint analysis of supernovas, BAO and CMB.

This would rule out the late-time vacuum decay in the context of the
present model, but the homogeneity of the interacting cosmological
term is an {\it ad hoc} hypothesis, which should be relaxed before
taking a definite conclusion. If matter is perturbed, the
interacting dark energy must also be, and we have to verify whether
such a perturbation is negligible or not. That is what we do in the
present paper. By writing our ansatz for the variation of vacuum
density in a covariant form, we derive a natural expression for the
linear perturbations in the cosmological term, which can be
integrated together with the relativistic equations for
perturbations in matter and radiation. We then construct the
predicted mass power spectrum, comparing with the observational
data. Our conclusion is that, assuming a scale invariant primordial
spectrum for the vacuum perturbations, a late-time vacuum decay
linear with $H$ is clearly ruled out also in this case.

The paper is organized as follows. In the next section we briefly
review the main features of the present model and its accordance
with other kind of observations. In Section III we obtain and
integrate the set of coupled perturbations equations for the vacuum,
matter, radiation and metric, obtaining the corresponding power
spectrum. In Section IV we outline our conclusions.

\section{The model}

There are two main motivations to consider that $\Lambda$ should
vary with time. The first one concerns the theoretically predicted
value for the cosmological constant, seen as a manifestation of
vacuum energy, and its observational value, a discrepancy that
mounts to 120 orders of magnitude. Considering some special
ingredients, like supersymmetry, this discrepancy can be reduced to
about 60 orders of magnitude, what is still a huge value. In view of
this problem, it would be interesting to have a mechanism that could
reduce the value of $\Lambda$, in order it could fill its role in
the inflationary era, attaining later its present, small value. This
could be achieved by allowing $\Lambda$ to vary with time. The
second motivation concerns the coincidence problem: the observed
value of the cosmological constant is of the same order of magnitude
of the ordinary matter density. Since the energy density of
$\Lambda$ is, in principle, constant and the energy density of
ordinary matter varies with time, this equivalence can only be
obtained in a given moment in the history of the evolution of the
universe. It is quite remarkable that this occurs exactly today,
when the process of galaxy formation is already essentially
finished. If we do not want to invoke a rather controversial
principle, like the anthropic one, it would be necessary to obtain a
mechanism that generate this fact. Giving dynamics to $\Lambda$ is a
natural approach to this coincidence problem. In any case, due to
energy conservation, the variation in time is equivalent to allow
$\Lambda$ to decay, generating matter, for example.
\par
The decay of $\Lambda$ into matter may solve the problems described
above and, at the same time, may somehow explain the origin of dark
matter. How to achieve such decaying process? The most simple idea
is to consider the conservation law
\begin{equation}
\dot\rho + 3\frac{\dot a}{a}(\rho + p) = 0,
\end{equation}
and suppose that $\rho = \rho_m + \rho_\Lambda$ and $p = p_m +
p_\Lambda$, where $m$ stands for {\it matter}. Imposing now that
$p_m = 0$ and $p_\Lambda = - \rho_\Lambda$, the above conservation
equation becomes
\begin{equation}
\dot\rho_m + 3\frac{\dot a}{a}\rho_m = - \dot\rho_\Lambda.
\end{equation}
As far as $\rho_\Lambda$ decays ($\dot\rho_\Lambda < 0$), the energy
density of the cosmological constant is converted into matter.
\par
This simple phenomenological approach faces a major drawback: In
doing so, we remain with the same number of equations, but we have
added a new function to be determined. Hence, we need an expression
which says how $\Lambda$ decay. One possibility is to invoke the
holographic principle. It says that the physical content of a given
system, defined in a volume $V$, is encoded in the area of the
surface enveloping this volume, $A$. This principle is motivated by
the black-hole thermodynamics: the entropy of a black-hole is given
just by the area of the event horizon, $S_{BH} = \frac{A}{4}$.
Another motivation, somehow related to the previous one, is given by
the AdS/CFT correspondence. For a review, see \cite{bousso}. In
cosmology, the problem can be addressed by establishing a connection
between the infrared and the ultraviolate cut-off defined in the
universe. The ultraviolet cut-off may be given, for example, by the
Planck's length, while the infrared cut-off is usually given by the
Hubble's radius. Since the Hubble radius is a function of time,
$\Lambda$ becomes a function of time. In reference \cite{Holografia}
the holographic principle is used in order to give dynamics to
$\Lambda$, but exploiting also the possibility that the infrared
cut-off is given by the particle horizon or the future event
horizon, besides the Hubble radius. In this way the authors conclude
that $\Lambda$ should behave as the square of the Hubble function,
and not linearly as for our model. The use of different infrared
cut-offs in \cite{Holografia} leads to different effective equations
of state for the mixture matter-$\Lambda$.
\par
Another possibility to determine the time variation of $\Lambda$ is
to consider quantum effects due to matter fields in the universe. In
this case, $\Lambda$ can be seen as a running parameter: the
renormalization group related to quantum fields in the dynamic
background of the FRW space-time implies that $\Lambda$ must run,
that is, must vary with time. In reference \cite{shapiro} (see also
\cite{Fabris}) this mechanism has been investigated and the authors
found that $\Lambda$ must behave generally as $\Lambda = \Lambda_0 +
\sigma H^2$, where $\Lambda_0$ is a constant and $\sigma$ is a parameter which depends on the ratio of the mass of the quantum fields with respect to Planck's mass, and on their nature (fermions or
bosons). This approach is quite
distinct from the approach of the present paper and from that of
reference \cite{Holografia}, since it is based on the effective
action due to quantum effects in the universe. A variation of such
model is the so-called $\Lambda XCDM$ model \cite{Grande}, where the
cosmological term is supposed to interact with a new field, called
{\it cosmon}, which has an equation of state $p_X = \omega_X\rho_X$.
\par
There is in the literature a large number of proposals leading to a
varying cosmological term. For some other frameworks, see
\cite{boussobis}. We will return later to the constraints due to the
LSS observational test on the models described above. Now, let us
describe in more details the model studied in this paper.

In the presence of pressureless matter and a time-dependent
cosmological term, the Friedmann equations in the spatially flat
case can be written as (we are using $8\pi G = c = 1$)
\begin{eqnarray}
\dot{\rho}_m + 3H\rho_m = -\dot{\Lambda}, \label{continuidade}\\
\rho_m + \Lambda = 3 H^2, \label{Friedmann}
\end{eqnarray}
where the dot means derivative with respect to the cosmological time
$t$, and $\rho_m$ is the matter density.

Let us take our late-time ansatz $\Lambda = \sigma H$, with $\sigma$
constant and positive. From the above equations we obtain the
evolution equation
\begin{equation} \label{evolucao}
2\dot{H} + 3 H^2 - \sigma H = 0.
\end{equation}
The solution, for $\rho_m, H > 0$, is given by \cite{Borges}
\begin{equation} \label{a}
a = C \left[\exp\left(\sigma t/2\right) - 1\right]^{2/3},
\end{equation}
where $a$ is the scale factor and $C$ is an integration constant
(another integration constant was taken equal to zero in order to
have $a = 0$ for $t = 0$). Taking the limit of early times, we have
$a \propto t^{2/3}$, as in the Einstein-de Sitter solution. It is
also easy to see that, in the opposite limit $t \rightarrow \infty$,
(\ref{a}) tends to the de Sitter solution.

With the help of (\ref{a}), and by using $\Lambda = \sigma H$ and
$\rho_m = 3H^2 - \sigma H$, it is straightforward to derive the
matter and vacuum densities as functions of the scale factor. One
has
\begin{equation}\label{rhodust}
\rho_m = \frac{\sigma^2 C^3}{3a^3} + \frac{\sigma^2
C^{3/2}}{3a^{3/2}},
\end{equation}
\begin{equation}\label{Lambdadust}
\Lambda = \frac{\sigma^2}{3} + \frac{\sigma^2 C^{3/2}}{3a^{3/2}}.
\end{equation}
In these expressions, the first terms give the standard scaling of
matter (baryons included) and vacuum densities, being dominant in
the limits of early and very late times, respectively. The second
ones are owing to the process of matter production, being important
at an intermediate time scale.

From (\ref{a}) we can also derive the Hubble parameter as a function
of time. It is given by
\begin{equation} \label{H}
H = \frac{\sigma/3}{1-\exp(-\sigma t/2)}.
\end{equation}
Finally, with the help of (\ref{a}) and (\ref{H}) we can express $H$
as a function of the redshift $z$, which leads to
\begin{equation} \label{Hz}
H(z) = H_0 \left[1-\Omega_{m0}+\Omega_{m0}(z+1)^{3/2}\right].
\end{equation}
Here, $\Omega_{m0}\equiv \rho_{m0}/(3H_0^2)$ and $H_0$ are the
present values of the relative matter density and Hubble parameter,
respectively.

Note that expression (\ref{Hz}) is valid only for late times, when
radiation can be neglected. For higher redshifts we need an
appropriate extension of it. As discussed in \cite{Jailson2,I}, for
times when radiation is important, the cosmological term and the
matter production are negligible. Therefore, a very good
approximation can be achieved by simply adding a conserved radiation
term to the total energy density. In this way we obtain
\begin{equation}
\label{Hgeral} H(z) \approx H_0 \left\{ \left[1 - \Omega_{m0}
+\Omega_{m0} (1 + z)^{3/2}\right]^2 + \Omega_{R0} (1+z)^4
\right\}^{1/2},
\end{equation}
where $\Omega_{R0} = \rho_{R0}/(3H_0^2)$ is the relative radiation
density at present.

We have analysed the redshift-distance relation for type Ia
supernovas \cite{Jailson2}, obtaining data fits as good as with the
spatially flat $\Lambda$CDM model. With the Supernova Legacy Survey
(SNLS) \cite{SNLS} the best fit is given by $h=0.70 \pm 0.02$ and
$\Omega_{m0} = 0.32 \pm 0.05$ (with $2\sigma$), with a reduced
$\chi$-square $\chi^2_r = 1.01$ (here, $h \equiv
H_0/$($100$km/s.Mpc)). With the inclusion of baryonic acoustic
oscillations in the analysis these results remain practically
unaltered. On the other hand, a joint analysis of the Legacy Survey,
BAO and the position of the first peak of CMB anisotropies has led
to the concordance values $h=0.69 \pm 0.01$ and $\Omega_{m0} = 0.36
\pm 0.01$ (with $2\sigma$), with $\chi^2_r = 1.01$ \cite{Jailson2}.
Note that the concordance value of $\Omega_{m0}$ is above the
current $\Lambda$CDM value \cite{sanchez}. This is a feature of the present
model, and a discussion about its origin can be found in
\cite{Jailson2} and \cite{I}.

\section{The mass power spectrum}

As the cosmological term has non-zero pressure, the inclusion of its
perturbations requires a relativistic treatment, and the first step
is to put the variation law for $\Lambda$ in a covariant form. In
comoving observers, it is possible to rewrite our late-time ansatz
$\Lambda = \sigma H$ as
\begin{equation} \label{ansatz_covariante}
\Lambda = \frac{\sigma}{3} u^{\nu}_{;\nu},
\end{equation}
where $u^{\nu}_{;\nu}$ is the covariant divergence of the cosmic
fluid $4$-velocity. Of course, this is not the only option to express
the ansatz covariantly. But it seems the most natural and
the simplest one.

We can now perturb this ansatz. By defining $\theta =
\partial_i \delta u^i$ and introducing the metric perturbation $h =
h_{kk}/a^2$, we obtain
\begin{equation} \label{ansatz_perturbado}
\delta \Lambda = \frac{\sigma}{3} \left(\theta -
\frac{\dot{h}}{2}\right).
\end{equation}

The other perturbation equations can be obtained by perturbing the
Einstein and the covariant energy conservation equations. This was
done in reference \cite{I}, where we consider conserved radiation,
matter and the interacting vacuum term as the energy components.
Here we will introduce two basic differences. The first one was
already discussed, namely the perturbation of the vacuum component.
Secondly, we will consider baryons independently conserved, with a
separated continuity equation, since they are not produced as vacuum
decays. This second novelty does not lead to important differences
in the resulting spectrum, but turns the analysis more precise. We
will also consider, as in reference \cite{I}, baryons decoupled from
radiation, a simplification which does not affect very much our
results. Indeed, in the case of the standard $\Lambda$CDM model such
a simplification leads to a difference about $10\%$ in comparison
with the exact analysis \cite{winfried}. Finally, we will suppose
that at any time the produced dark particles have the same velocity
field of the pre-existing interacting fluid formed by dark matter
and vacuum. This is a reasonable hypothesis, since we are dealing
with matter production in the low energy limit at large times.

On this basis it is straightforward to derive, in the synchronous gauge, the set of equations
\begin{eqnarray}
\ddot{h} + 2H\dot{h} &=& \rho_{dm} \delta_{dm} + \rho_b \delta_b +
2\rho_R \delta_R - 2\Lambda \delta_{\Lambda},\\
\dot{\delta}_R + \frac{4}{3} \left( \frac{v_R}{a} -
\frac{\dot{h}}{2} \right) &=& 0,\\
\dot{v}_R &=& \frac{k^2}{4a} \delta_R,\\
\dot{\delta}_{dm} - \frac{\dot{\Lambda}}{\rho_{dm}} \delta_{dm} +
\frac{v_{dm}}{a} - \frac{\dot{h}}{2} &=& -
\frac{\dot{\Lambda}}{\rho_{dm}} \delta_{\Lambda} -
\frac{\Lambda}{\rho_{dm}} \dot{\delta}_{\Lambda},\\
\dot{v}_{dm} + \left( \frac{\dot{\rho}_{dm}}{\rho_{dm}} + 4H \right)
v_{dm} &=& - \frac{k^2\Lambda}{a\rho_{dm}} \delta_{\Lambda},\\
\dot{\delta}_b - \frac{\dot{h}}{2} &=& 0 \label{17}.
\end{eqnarray}
In these equations $k$ is the wave number; $\rho_{dm}$ and
$\rho_{b}$ are the energy densities of dark matter and baryons,
respectively; $\delta_i = \delta \rho_i/\rho_i$ defines the density
contrast of each component; $v_{dm} = a\, \theta$ and $v_R$ are the
peculiar velocities of dark matter and radiation, respectively. As
the baryonic component is pressureless and independently conserved,
its peculiar velocity remains uncoupled and tends to zero, being
then neglected.

Therefore, for a given background, we have a system of six equations
with seven variables, that can be reduced to a system of five
equations with six variables if we use $\dot{\delta}_b = \frac{{\dot
h}}{2}$ (equation (\ref{17})). To solve the resulting system of five
equations, it is also necessary to add our previous equation
(\ref{ansatz_perturbado}). In this way we can, for example,
eliminate $v_{dm}$ from the system. Using our background solution
(see Section II), changing the independent variable from the
cosmological time $t$ (after making it dimensionless by redefining
$H_0 t \rightarrow t$) to the scale factor $a$, and fixing $a_0
= 1$, we finally obtain the system
\begin{eqnarray}
\delta_b''+\left(\frac{g}{f^2}+\frac{2}{a}\right)\delta_b' &=&
\frac{3}{2f^2}\left(2\Omega_R \delta_R + \Omega_b \delta_b + \Omega_m
\delta_m - 2\Omega_{\Lambda} \delta_{\Lambda}\right),\\
\delta'_R + \frac{4}{3} \left(\frac{v_R}{af}-\delta_b'\right) &=& 0,\\
v_R' - \frac{k^2}{4af}\delta_R &=& 0,\\
\delta'_m - \frac{1}{1+r}\left(r'-\frac{3r}{a}\right)\delta_m &=&
-\left[ \frac{1}{1+r}\left(r'-
\frac{3r}{a}\right)+\frac{3}{a}\right]\delta_{\Lambda}-r\delta'_{\Lambda},\\
\delta'_{\Lambda}+\left\{\frac{f'}{f}+\left[\frac{1+4r}{(1+r)a}-\frac{r'}
{1+r}\right] + \frac{k^2r}{3af^2}\right\} &\delta_{\Lambda}& =
\nonumber\\ &-&
\frac{a}{3}\left\{\delta_b''+\left[\frac{f'}{f}+\frac{2+5r}{(1+r)a}-\frac{r'}{1+r}\right]\delta_b'\right\},
\end{eqnarray}
where the prime means derivative with respect to $a$. Here we are
using the definitions
\begin{eqnarray}
\Omega_R &=& \frac{\Omega_{R0}}{a^4},\\
\Omega_b &=& \frac{\Omega_{b0}}{a^3},\\
\Omega_{dm} &=&
\frac{1}{a^3}\left(\Omega_{dm0}-\Omega_{\Lambda0}+\Omega_{\Lambda0}^2\right)
+ \frac{1}{a^{3/2}}\left(\Omega_{\Lambda0}-\Omega_{\Lambda0}^2\right),\\
\Omega_{\Lambda} &=&
\Omega_{\Lambda0}^2+\frac{1}{a^{3/2}}\left(\Omega_{\Lambda0}-\Omega_{\Lambda0}^2\right),\\
r &=& \frac{\Omega_{\Lambda}}{\Omega_{dm}},\\
g &=& a \left( - \Omega_R -
\frac{\Omega_b}{2}-\frac{\Omega_{dm}}{2}+\Omega_{\Lambda}\right),\\
f &=& a\left(\Omega_R + \Omega_b + \Omega_{dm} +
\Omega_{\Lambda}\right)^{1/2},
\end{eqnarray}
with $\Omega_{i0}=\rho_{i0}/(3H_0^2)$ meaning, as before, the
present relative density of each component. Note that we are doing
the same approximation used in (\ref{Hgeral}), i.e., we are taking
$\Omega_{b0} + \Omega_{dm0} + \Omega_{\Lambda0} = 1$, since
$\Omega_{R0} \approx 8 \times 10^{-5}$ is negligible as compared to
the other relative densities.

The system above can now be numerically integrated with appropriate
initial conditions. In fixing the initial conditions there is a
possible difficulty due to the fact that, in the present model, the
cosmological term does not reduce to a constant $\Lambda$ for large
redshifts. Hence, strictly speaking we should solve the
Einstein-Boltzmann system for perturbed quantities. As discussed
above, however, we can consider simply the system with a radiative
fluid, baryons, dark matter and the cosmological term from very high
redshifts (typically up to $z = 10^{12}$). In the case of the
$\Lambda$CDM model, the results differ from a more exact analysis
for very large scales by some values of the order of $10\%$. For
these large scale perturbations there are some problems with
statistical variance. But for scales where the linear approximation
is good enough and there is no variance problem, the agreement is
quite reasonable. In avoiding integrating the complete Einstein-Boltzmann system,
there is an inevitable discrepancy between the
evaluated spectrum and the real one (which is of the order of $10\%$ as said above).
Hence, a small correction must be added. We introduce this correction (essentially a
$k$-dependent factor) using as a reference system the $\Lambda$CDM model and
the BBKS transfer function \cite{jerome}. This is an improvement with respect to the
method employed in \cite{I}. But, even if such correction
is not introduced, the final conclusions remain the same.

We can compare our theoretical results with two different sets of
observational data, those coming from the 2dFGRS \cite{2dFGRS} or
the SDSS \cite{SDSS} programs. In the present work we will restrict
ourselves to the 2dFGRS data, and this for one reason: they cover a
small range of scales ($0.01\,Mpc^{-1} < k h^{-1} <
0.185\,Mpc^{-1}$), and for most of the data the linear approximation
is quite good, and moreover the error bars are small. The SDSS data
cover values of $kh^{-1}$ up to $0.3\,Mpc^{-1}$, and we must worry
about non-linearity effects. In particular, using the SDSS data we
can hardly avoid the use of the covariance matrix (which is the case
also for the 2dFGRS data for $kh^{-1} > 0.15\,Mpc^{-1}$, strictly
speaking). Since we intend to stay at the regime of validity of the
linear approximation, the 2dFGRS set seems more convenient. We
remark {\it en passant} that there are some claims in the literature
of the incompatibility of parameter estimations using the 2dFGRS or
SDSS data \cite{cole}. We will restrict ourselves to the 2dFGRS data
with $0.02\, Mpc^{-1} < kh^{-1} < 0.15\, Mpc^{-1}$, to avoid
problems with uncertainties and non-linearity \cite{percival}.

We will consider that the vacuum perturbations present a
scale-invariant spectrum for very high redshifts, with the same
amplitudes for the matter perturbations. In other words, we take
$\delta_{\Lambda} = \delta_{dm} = \sqrt{k}$ for, say, $a =
10^{-12}$. After integration, the square of $\delta_{dm}(k)$ gives
the mass power spectrum up to a normalization factor. In order to
normalize it we use the BBKS transfer function \cite{jerome}, where
the CMB results are used to normalize the spectrum, giving the
correct spectrum for the spatially flat standard model.

\begin{center}
\begin{figure}[]
\begin{minipage}[t]{0.25\linewidth}
\includegraphics[width=\linewidth]{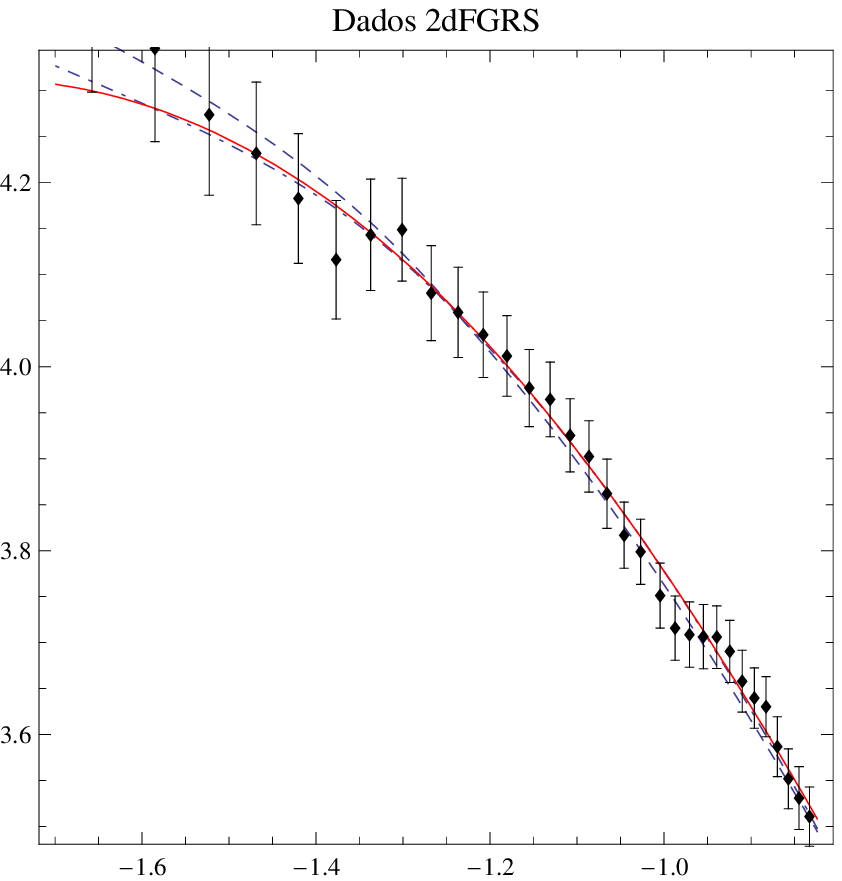}
\end{minipage} \hfill
\begin{minipage}[t]{0.25\linewidth}
\includegraphics[width=\linewidth]{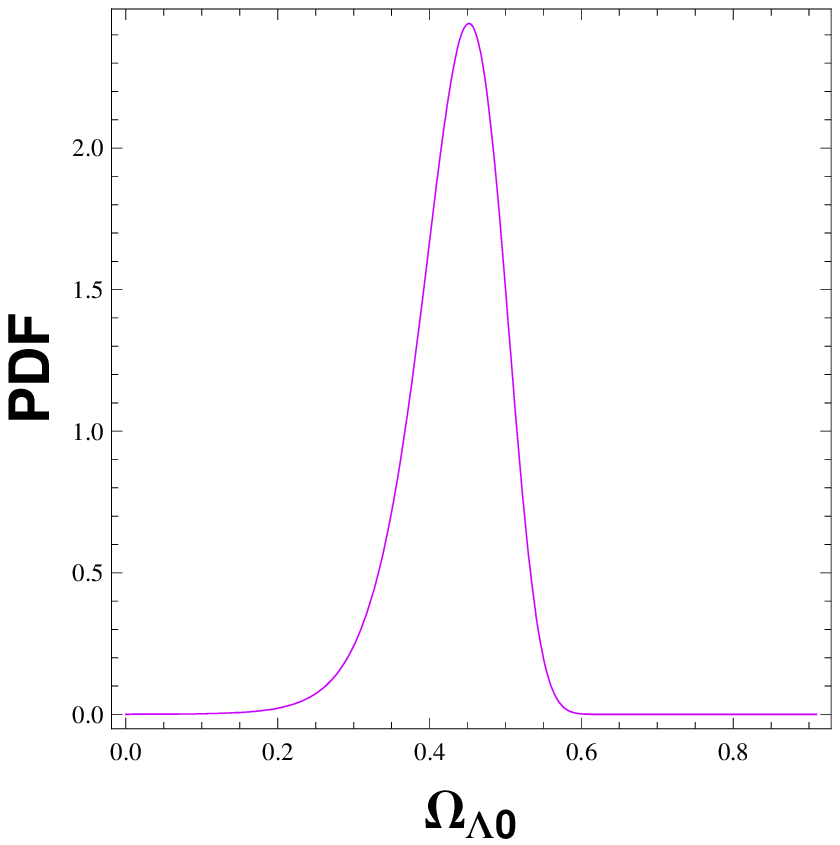}
\end{minipage} \hfill
\begin{minipage}[t]{0.25\linewidth}
\includegraphics[width=\linewidth]{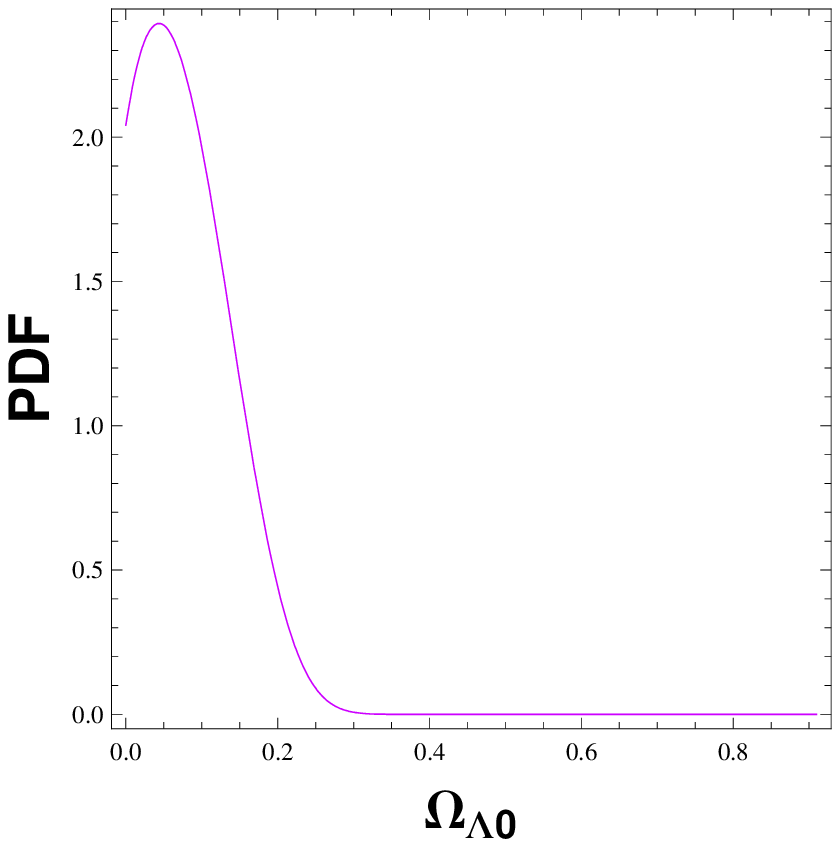}
\end{minipage} \hfill
\caption{{\protect\footnotesize The best fit models for the
$\Lambda$CDM model with $\Omega_{dm0} = 0.24$ (continuous line), the
decaying $\Lambda$ model without the perturbation of the $\Lambda$
term (dashed line) and with the perturbation in the $\Lambda$ term
(dot-dashed line), when conserved baryons are not included. At the
center, it is shown the PDF distribution of the dark energy density
parameter when the cosmological term is not perturbed and at right
the same for the case the cosmological term is perturbed.}}
\end{figure}
\end{center}

\begin{center}
\begin{figure}[]
\begin{minipage}[t]{0.25\linewidth}
\includegraphics[width=\linewidth]{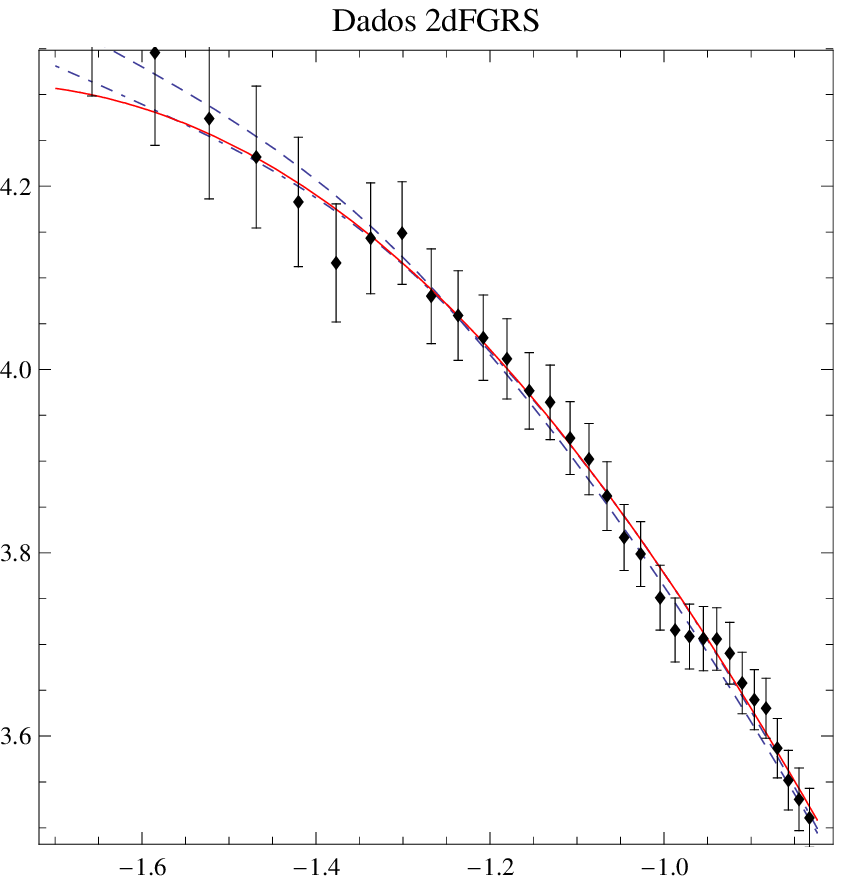}
\end{minipage} \hfill
\begin{minipage}[t]{0.25\linewidth}
\includegraphics[width=\linewidth]{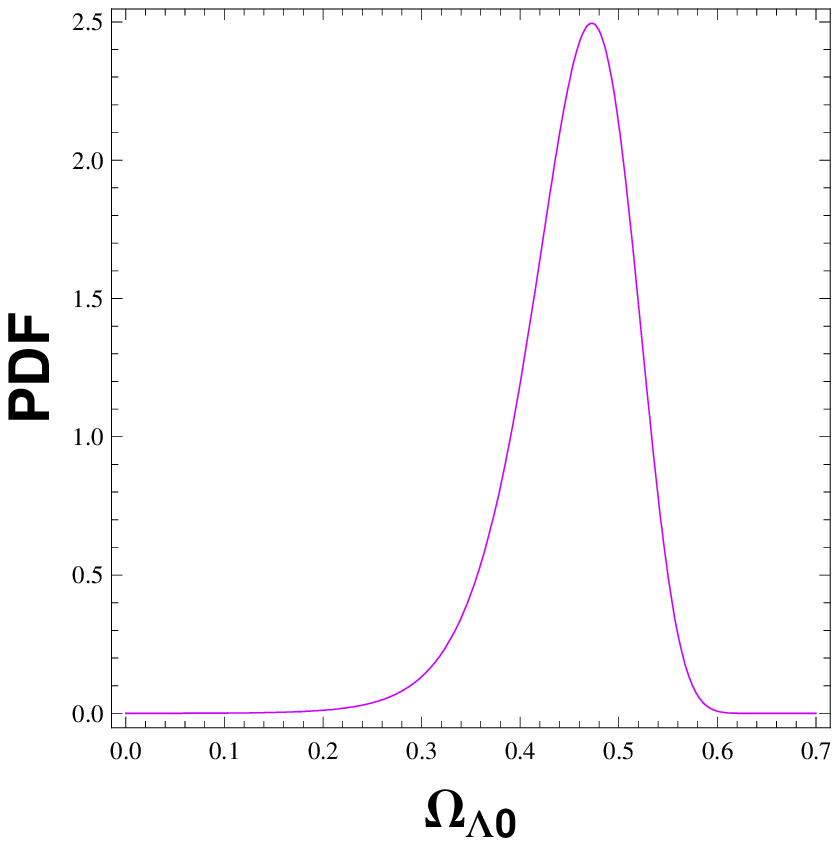}
\end{minipage} \hfill
\begin{minipage}[t]{0.25\linewidth}
\includegraphics[width=\linewidth]{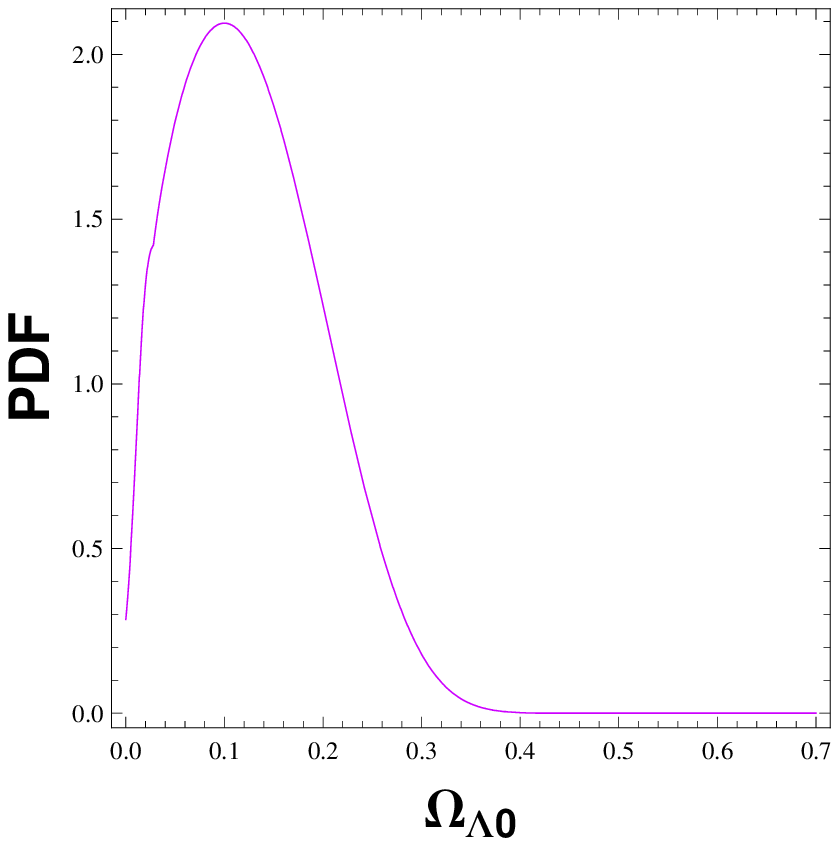}
\end{minipage} \hfill
\caption{{\protect\footnotesize The best fit models for the
$\Lambda$CDM model with $\Omega_{dm0} = 0.24$ (continuous line), the
decaying $\Lambda$ model without the perturbation of the $\Lambda$
term (dashed line) and with the perturbation in the $\Lambda$ term
(dot-dashed line), when conserved baryons are included. At the
center, it is shown the PDF distribution of the dark energy density
parameter when the cosmological term is not perturbed and at right
the same for the case the cosmological term is perturbed.}}
\end{figure}
\end{center}

Restricting ourselves to the linear regime, we can use confidently
the $\chi^2$ statistics, defining the statistic fitness parameter
\begin{equation}
\chi^2 = \sum_i\biggr(\frac{{\cal P}_i^o - {\cal
P}_i^t}{\sigma_i}\biggl)^2,
\end{equation}
where ${\cal P}_i^o$ is the observational data for the $i\,th$ value
of $k$, $\sigma_i$ its observational error bar and ${\cal P}_i^t$
the corresponding theoretical value. This parameter depends on the
densities, and from it a probability distribution can be
constructed, by defining
\begin{equation}
F = A\exp(-\chi^2/2),
\end{equation}
$A$ being a normalization constant.

Four situations are considered here, combining the inclusion or not
of a baryonic component which is conserved separately, and the
possibility that the cosmological term is perturbed or not. In
Figure $1$ we exploit the case where conserved baryons are not
included, which means to do $\Omega_{b0} = 0$ in our system of
perturbed equations. The best fitting for the $\Lambda$CDM model
(with $\Omega_{dm0} = 0.24$ \cite{cole}), for the case the
cosmological term is not perturbed (see \cite{I}) and for the case
it is perturbed are shown, together with the corresponding
probability distribution function (PDF) from the $\chi^2$ analysis,
both with a perturbation in the cosmological term and without it.
The same graphics are displayed in Figure $2$, but considering the
inclusion of conserved baryons, with $\Omega_{b0} = 0.044$. It is
easy to verify that the inclusion of the perturbation of the
cosmological term displaces strongly the PDF for dark energy to the
left: less dark energy is necessary, and consequently more dark
matter. As an example, when conserved baryons are included, the best fitting models
ask for $\Omega_{\Lambda0} = 0.47$ ($\Omega_{dm0} = 0.49$) when
the cosmological term is not perturbed, and $\Omega_{\Lambda0} = 0.10$
($\Omega_{dm0} = 0.86$) when $\Lambda$ is perturbed.
The best fits are also shown in the table below.

\begin{center}
\begin{tabular}{|c|c|c|c|c|}
\hline &$\delta\Lambda \neq 0$, $\Omega_{b0} = 0$&$\delta\Lambda =
0$, $\Omega_{b0} = 0$
&$\delta\Lambda \neq 0$, $\Omega_{b0} = 0.04$&$\delta\Lambda = 0$, $\Omega_{b0} = 0.04$\\
\hline $\chi^2_{r}$&$0.31$&$0.35$&$0.31$&$0.35$\\
\hline
$\Omega_{dm0}$&$0.96$&$0.55$&$0.86$ &$0.49$\\
\hline
$\Omega_{\Lambda0}$&$0.04$&$0.45$&$0.10$&$0.47$\\
\hline
\end{tabular}
\end{center}

We can now compare our results with the other models of varying
$\Lambda$ described in section II. In reference \cite{Fabris} a full
comparison of the theoretical results with the 2dFGRS data has been
made, considering perturbations also in the cosmological term. In
this model there is a free parameter represented by the constant
$\sigma$. It was found that there is a very good agreement of the
theoretical results with the observational data if $|\sigma| \leq 10^{-4}$. 
This implies that the mass of the quantum field must not
be large compared with the Planck's mass. Hence, the resulting
scenario is very similar to the $\Lambda CDM$ model, which is
re-obtained when $\sigma \rightarrow 0$ (no running of the
cosmological constant). These results are in agreement with those
of reference \cite{pavon}, where a dark matter/dark energy
interacting model has been studied using also matter power spectrum
data: the agreement between theory and observation is assured when
the interaction is very weak, that is, $\Lambda$ is a slow-varying
function of time. The present model, on the other hand, does
not have any free parameter that could connect it, in some limit, to
the $\Lambda CDM$ model.
\par
In reference \cite{Grande} the $\Lambda XCDM$ model has been studied
using also the growth of linear perturbations, and restrictions on
the parameter space of the model were obtained. But the cosmological
term was considered as smooth: the authors argue that the influence
of the perturbations in $\Lambda$ is not significative for modes
well inside the Hubble horizon, those concerned by the 2dFGRS data.
A full comparison with the matter power spectra data for the
$\Lambda XCDM$ model has been performed in \cite{solabis}. There are
two free parameters in the model, the parameter $\sigma$ which has
the same meaning as in the running cosmological model of reference
\cite{Fabris}, and the equation of state parameter $\omega_X$ of the
cosmon component, which exchange energy with the cosmological term.
For a given region of this bidimensional parameter space the
agreement with the data is excellent. Again, this happens near the
cosmological constant case, which is a particular limit of those
parameters.
\par
The model of reference \cite{Holografia} also implies, as we have
seen, a variation of $\Lambda$ proportional to $H^2$. The nature of
the infrared cut-off (Hubble radius, particle horizon or future
event horizon) changes the effective equation of state, but not the
quadratic dependence of $\rho_\Lambda$ on $H$. Hence, we can expect
that the results using the matter power spectrum constraints should
be similar to those found in reference \cite{Fabris}. In their work,
the authors of reference \cite{Holografia} have used a Newtonian
approach, imposing that the cosmological term remains a smooth
component, not being perturbed. Their main results indicate that
there are growing modes only when the effective equation of state
implies that the energy conditions are not violated. An
investigation similar to that made in the present work or in
reference \cite{Fabris}, considering a full relativistic approach,
perturbing also the cosmological term and using extensively the
2dFGRS data, may be relevant for the case of reference
\cite{Holografia}.
\par
In all these cases, we can remark that the models for which
$\Lambda$ varies with $H^2$ contain at least one free parameter
assuring that the $\Lambda CDM$ limit is contained in the model.
This is not our case. Moreover, since $H$ is a small number, in the
former models the variation of the cosmological term is small. A
linear dependence, as we have assumed, containing no $\Lambda CDM$
limit, implies comparatively a high interaction, which seems to be
ruled out by observations, in agreement with the results of
reference \cite{pavon}.

\section{Conclusions}

From a theoretical point of view, the hypothesis of vacuum decay is
interesting in several aspects. First of all, it may naturally
conciliate the observed cosmological term with a huge initial value
for the vacuum density through a relaxing mechanism in the expanding
space-time. Secondly, in the realm of our thermodynamic ansatz, we
have a non-singular and inflationary early phase in the universe
evolution, with the presently observed matter content generated by a
primordial vacuum transition \cite{SC2}.

A late-time vacuum decay depends on the mass of the produced
particles. This possibility would be interesting as an
explanation for the cosmic coincidence, since the suppression of the
matter contrast owing to matter production begins to have importance
when the cosmological term starts to dominate the cosmic expansion
\cite{I}. Until now it has also been survived to a precise joint
analysis of supernovas, BAO and CMB observations, with a good
concordance for the two free model parameters, i.e., the present
values of the Hubble constant and of the relative matter density
\cite{Jailson2}.

In this paper we have extended a previous study of structure
formation in this context \cite{I}, investigating the consequences
of matter production for the mass power spectrum. We had already
shown that, in the case of a strictly homogeneous cosmological term,
the suppression in the matter contrast leads to a disagreement with
the observed spectrum, unless the matter density parameter is as
high as $\Omega_{m0} = 0.48$ (for which the accordance is
excellent). This value is outside the current dynamical limits on
$\Omega_{m0}$ and above the concordance value obtained from the
joint analysis of supernovas, BAO and CMB, $\Omega_{m0} = 0.36$
\cite{Jailson2}.

Now we have considered the possibility of perturbations in the
cosmological term, a more natural assumption. Assuming that such a
term presents the same scale-invariant primordial spectrum as dark matter, we
conclude that the agreement with the observed spectrum, for any
acceptable value of $\Omega_{m0}$, is even worse. We have also considered
the case in which the initial vacuum perturbations are zero, but the fitting with
observations is still bad, unless for a very high matter density.
This seems to rule out a late-time running of the vacuum term, at least in the recipe
of the present model.

Of course, one could test other possible amplitudes and forms for
the vacuum primordial spectrum (despite its unnaturalness). In fact, we
have tested a large range of amplitudes, but the obtained power
spectrums are not significantly better. We have also tested values
for the present mass density of decoupled particles above the
baryonic mass density, supposing the presence of a massive dark
particle aside the light one produced by the vacuum decay. Also in
this case the fitting between the observed and predicted power
spectrums is poor.

Let us remind, however, that discarding a late-time
variation of the cosmological term in the present model does not
mean to discard the general idea of vacuum decay.
If dark energy is
a manifestation of quantum vacuum in the curved, expanding
background, the inclusion of a decaying cosmological term in
Einstein equations is as natural as the inclusion of a genuine
cosmological constant. In the realm of the present model, the
disagreement with large structure observations indicates that the
cosmological term does not vary at late times. Nevertheless, a
different ansatz for the vacuum variation may lead to a better
accordance.\footnote{Another possible scape for this situation would
be supposing that our ansatz is not a good approximation for the
present universe, because it is still not a quasi-de Sitter
space-time. In this case, however, it would be surprising that a
genuine cosmological constant fits so well the observations.}

Even if the vacuum decay is restricted to very early times - due to
the masses of the produced particles - we have interesting
consequences, as already discussed. A precise comparison of such a
scenario with observations, in the context of our thermodynamic
ansatz \cite{SC2}, is still in order. Particularly, its necessary to
find the primordial spectrum of matter perturbations generated in
the inflationary phase of the model, to be tested against the
observed spectrum of anisotropies in the CMB. This research is
already in progress.

\section*{Acknowledgements}
H.A.B. is supported by CAPES. S.C. is partially supported by CAPES
and CNPq, and he is thankful to Prof. Reza Tavakol for the warm
hospitality in Queen Mary, University of London. J.C.F. is partially
supported by CNPq, FAPES and the Brazilian-French scientific
cooperation CAPES/COFECUB, and he thanks {\it GR$\epsilon$CO}, IAP,
France, for the kind hospitality.


\end{document}